\documentclass[proof]{WileyASNA-v1}

\articletype{Article Type}%

\received{26 April 2016}
\revised{6 June 2016}
\accepted{6 June 2016}

\begin{document}

\title{Discovering periodic and repeating nuclear transients in the XMM-Newton archives}

\author[1]{Natalie A. Webb*}

\author[1]{Vincent Foustoul}
\author[1]{Robbie Webbe}

\author[2]{Matteo Bachetti}
\author[3]{Erwan Quintin}
\author[4]{Laurent Michel}

\authormark{Webb \textsc{et al}}

\address[1]{\orgdiv{Universit\'e de Toulouse, CNRS, CNES}, \orgname{Institut de Recherche en Astrophysique et Plan\'etologie}, \orgaddress{\state{Toulouse}, \country{France}}}

\address[2]{\orgdiv{INAF}, \orgname{Osservatorio Astronomico di Cagliari}, \orgaddress{\state{via della Scienza 5 I-09047 Selargius (CA)}, \country{Italy}}}

\address[3]{\orgdiv{European Space Astronomy Centre
(ESAC)}, \orgname{European Space Agency(ESA)}, \orgaddress{\state{Madrid}, \country{Spain}}}

\address[4]{\orgdiv{CNRS-INSU}, \orgname{Observatoire Astronomique de Strasbourg}, \orgaddress{\state{Strasbourg}, \country{France}}}

\corres{*\email{Natalie.Webb@irap.omp.eu}}

\abstract{The regions around massive black holes can show X-ray variability on timescales from seconds to decades. Observing many black holes over different timescales can enhance our chances of detecting variability coming from (partial) tidal disruption events, massive black hole binaries, changing state AGN, blazar activity and much more. X-ray catalogues with hundreds of thousands of detections are treasure troves of such sources, which require innovative methods to identify these black holes.  We present the current XMM-Newton catalogues available and describe several examples of tidal disruption events (TDEs) and quasi-periodic eruption sources that have been found whilst mining this data. We describe preliminary work on a search for periodic variables in the XMM-Newton EPIC archival data, with the example of finding new massive black hole binaries. We also describe the STONKS pipeline that is now in the XMM-Newton automatic reduction pipeline and the near real-time alert system that allows the follow-up of new and fading transients. We provide examples of fading sources that are newly identified candidate TDEs. 
}




\maketitle


\section{Introduction}\label{sec:intro}
\label{sec:intro}

Observing the X-ray sky enables us to detect the hottest and most energetic phenomena in the Universe. At such short wavelengths, we observe matter accreted onto massive black holes in the centres of galaxies or onto stellar mass black holes in binaries called X-ray binaries. Such wavelengths allow us to identify stellar flares from active stars, supernova remnants, neutron stars, white dwarfs, galaxy clusters and even aurora on planets or comets. Modern X-ray catalogues contain hundreds of thousands of sources, some of which have data taken over more than two and a half decades. Combining several catalogues can provide data for some sources that span almost forty years, and for which there can be hundreds of observations. These are rich resources with which to search for variability and transients in X-ray bright objects. 

 {\it XMM-Newton}, a {\it European Space Agency} X-ray observatory \citep{jans01} and the largest X-ray telescope to be launched to date, has been observing the X-ray, ultra-violet and optical sky for 25 years. The {\it XMM-Newton Survey Science Centre} \citep[XMM-SSC, ][]{wats01}, in collaboration with the {\it XMM-Newton} Science Operations Centre (SOC) have put together catalogues of all the detections and sources from pointed observations made with the {\it European Photon Imaging Cameras} (EPIC). The most recent version of the detection catalogue, 4XMM-DR14, was released in July 2024 and contains 1035832 detections from 692109 sources. To complement the 336 columns of data in this catalogue, spectra and lightcurves are provided for the brightest 36\% of detections, for which there are at least 100 X-ray counts. 8380 sources are qualified as variable within an observation, at the 5$\sigma$ level and some sources have as many as 90 detections over the last 25 years \citep{webb20}. The median fluxes are $\sim$5.2$\times$10$^{-15}$ erg cm$^{-2}$ s$^{-1}$ (0.2-2.0 keV) and $\sim$1.2$\times$10$^{-14}$ erg cm$^{-2}$ s$^{-1}$ (2-12 keV). The XMM-SSC provides the results of cross-correlating the detections with 222 multi-wavelength catalogues. The XMM-SSC also produces a catalogue of sources produced by stacking overlapping observations, 4XMM-DR14s \citep{traul20}, to achieve better sensitivity and improve source parameters, as well as provide direct access to long-term flux variability. Some sources have a total of ~3 Ms of observations. 4XMM-DR14s is built from 1751 groups drawn from 10332 observations and contains 427524 sources, of which 329972 have several contributing observations, with a total of 1.8 million individual flux measurements. The median source fluxes are $\sim$2.5$\times$10$^{-15}$ erg cm$^{-2}$ s$^{-1}$ (0.2-2.0 keV) and $\sim$6.8$\times$10$^{-15}$ erg cm$^{-2}$ s$^{-1}$ (2.0-12.0 keV). A third X-ray catalogue provides all of the detections made during the telescope slews between observations. The EPIC cameras take data during the slew, configured in the observing mode set in the previous pointed observation and with the Medium filter in place. The slew speed is 90 degrees per hour, so a source is usually observed for about 20s, but due to the long MOS frame time (2.6 s), the sources are elongated and can not be exploited. However, the pn data can be exploited and a catalogue of these detections has had three releases. The most recent version of the slew catalogue \citep{saxt08} (XMMSL3) was released by the XMM-SSC on 19th February 2025. This version contains data taken from August 2001 to August 2023, thus almost 10 years of extra data compared to XMMSL2. 140735 detections which relate to 116598 unique sources are included in the catalogue, where some sources are detected as many as 78 times. The advantage of this catalogue compared to the 4XMM-DR14(s) catalogues is that it covers 93.7\% of the sky, compared to only a few percent in the pointed data catalogues. The median detection flux is 2.7 $\times$ 10$^{-12}$ erg cm$^{-2}$ s$^{-1}$. Since release, a minor issue has been identified, where a few detections are duplicated. The list of duplicates can be found on the XMM-SSC website\footnote{\url{http://xmmssc.irap.omp.eu/}}. A fourth catalogue is also produced by the SOC, in collaboration with the SSC, for sources detected by the Optical Monitor \citep{maso01}. This is the Serendipitous Ultra-Violet Source Survey \citep[SUSS, ][]{page12}, where the most recent version, 6.2, comprises data from February 2000 to November 2022 for detections in the six filters, where available (UVW2, UVM2, UVW1, U, B, V). This version contains 9920390 detections, corresponding to 6659554 unique sources, of which 1225117 have multiple entries. The UV magnitudes have limiting values of around 21. The data in this catalogue are taken contemporarily with the X-ray pointed observations and therefore provide an extra dimension to the X-ray data.

\section{Nuclear transients found in the XMM-Newton catalogues}
\label{sec:transients}

There are many types of nuclear transients. Tidal disruption events (TDEs) occur when a star crosses the tidal radius of a massive black hole (MBH). At this radius, the tidal forces exceed the binding energy of the star, ripping it apart \citep{rees88}. About half of the stellar mass will be captured and accreted by the black hole on the order of $\sim$1 year \citep{hill75}. This results in an outburst with a peak luminosity of up to 10$^{45}$ erg s$^{-1}$ \citep[e.g.][]{stru09}. Some TDEs discovered through the {\it XMM-Newton} catalogues include : 2XMMi J184725.1-631724 in the galaxy IC 4765-f01-1504, situated at z=0.0353, with a fairly low black hole mass of 0.06 - 4 $\times$ 10$^6$ M$_\odot$ \citep{lin11}; 3XMM J152130.7+074916 \citep{lin15} at the centre of the dwarf galaxy SDSS J152130.72+074916.5 at z = 0.17901 (866 Mpc);  3XMM J215022.4$-$055108 \citep{lin18,lin20,chen18} reached a peak luminosity of $\sim$1 $\times$ 10$^{43}$ erg s$^{-1}$ (0.2-10.0 keV) with a possible mass range of 5.3-12 $\times$ 10$^4$ M$_\odot$, making it an excellent candidate for an intermediate mass black hole (IMBH); 3XMM J141711.1+522541 in the inactive S0 galaxy SDSS J141711.07+522540.8 at z = 0.418 \citep{lin16}. Supposing that the source reached the Eddington luminosity around the peak, the black hole mass would be $\sim$10$^5$ M$_\odot$ \citep{lin16}; the similar TDE SDSS J120136.02+300305.5 \citep{saxt12}; and XMMSL1 J074008.2-853927 \cite{saxt17}, which is notably different as the X-ray spectrum is dominated by a power law with $\Gamma\sim2$, rather than a soft blackbody.  More extreme still is the TDE 3XMM J150052.0+015452 found at z$\sim$0.145 \citep{lin17b}.  Modelling of the X-ray spectra indicates that the source was in a super-Eddington state for about five years, before dropping to approximately Eddington luminosity, where it is expected to remain for more than two decades \citep{lin22}.

The star may not, however, be totally disrupted if the star fails to cross the tidal radius. It is possible that only the outer layers of the star are ripped off, leaving the stellar core to reform as a star \citep{rees88}. Depending on the orbital parameters and the mass of the star, it will become unbound or orbit the MBH, undergoing partial disruption at periastron or intersect the accretion stream/disc, giving rise to outbursts \citep[e.g.][]{dai10}. Partial TDEs should be observed more frequently than complete TDEs as the rate of encounters scales with the pericentre distance and because there is more chance to observe the repeated bursts \citep{guil13}. One example discovered with {\it XMM-Newton} is 2XMM J011028.1-460421, commonly known as Hyper Luminous X-ray source 1 \citep[HLX-1, ][]{farr09}, which reaches a maximum luminosity of $\sim$1.3 $\times$ 10$^{42}$ erg s$^{-1}$ (0.2-10.0 keV) and has a mass of $\sim$10$^4$ M$_\odot$ \citep{gode12}, also making it an IMBH. It is highly variable, showing eight outbursts since 2008 each with a factor $\sim$50 rise in luminosity. The time between outbursts has varied significantly since the first pair of bursts, going from $\sim$379d, to $\sim$351d, to $\sim$350d, to $\sim$400d, to $\sim$490d and then $\sim$800d and it is now more than 2600d since the last eruption. Such an evolution can be expected from an intermediate mass ratio system in a very elliptical orbit (eccentricity close to unity), so the system behaves initially like an intermediate mass ratio inspiral (IMRI) and then becomes an intermediate mass ratio outspiral (IMRO)! Other partial tidal disruption events have been discovered in the X-ray, but with {\it eROSITA} e.g. eRASSt J045650.3-203750, which is extremely similar to HLX-1 \citep{liu24} or Swift J023017.0+283603 with {\it Swift} \citep{evan23}, which has shown around nine bursts.
 
Another transient feature has also been associated with some TDEs, quasi periodic eruptions \citep[QPEs,][]{mini19}. These are strong, quasi-periodic bursts of soft X-ray emission, detected as the source declines towards quiescence. The first QPE was identified by \cite{mini19}, when observing the long-term decline of a TDE discovered with the slew catalogue \citep{saxt11}.  They observed the X-ray count rate increase by up to two orders of magnitude over 1 h every 9 h in the Seyfert 2 galaxy GSN 069. Since then, a dozen other systems have shown QPEs including RX J1301.9+2747 \citep{gius20}, eRO-QPE1 and eRO-QPE2 \citep{arco21}, Tormund \citep{quin23} and tentatively XMMSL1 J024916.6-041244 \cite{chak21}. However, \cite{arco22} showed that the time between bursts is not always quasi-periodic, with bursts sometimes arriving in pairs. Further, the burst start and peak times vary for different energies.

Accreting massive black holes (Active Galactic Nuclei, AGN) can show variability due to the accretion processes, either through state changes, for instance through the disc instability model, or show flares or even change completely, what is called the {\it changing look AGN}, e.g. \cite{kara25}. Variability can also arise following the merger of two galaxies. With time, the two black holes will sink to the centre of mass and orbit each other due to the angular momentum remaining from the merger.  This creates what is called a dual AGN, if there is sufficient gas and dust available to be accreted onto the two black holes. The binary will then harden (lose angular momentum) through dynamical friction through stellar and dust interactions. During the shrinking of the orbital separation of the two MBH, their velocity increases and dynamical friction becomes less and less effective \citep{yu02} until gravitational wave emission starts to remove angular momentum from the binary.  This is known as a massive black hole binary. Eventually, the two black holes should merge, which may be an important process in the growth of supermassive black holes. Discovering these systems can help to constrain the role of this mechanism in massive black hole evolution. The binaries can be detected through their (quasi-)periodic electromagnetic emission, from the modulation of the accretion disc emission over the orbit or through an orbiting over-density or {\it lump} in the circumbinary disc, e.g. \cite{fous25}. In X-ray, this emission is expected to emanate from the discs (disc black body emission) from the individual black holes and/or the X-ray corona around the black hole(s), e.g. \cite{cocc24}, thus the X-ray emission comes from the central regions around the black holes. If the X-ray emission is from the disc, the black holes will not be very massive, as the supermassive black hole (10$^9$ - 10$^{10}$ M$_\odot$) discs radiate primarily in the optical and ultra-violet. Alternatively they can be detected using gravitational waves, either through the different pulsar timing arrays (PTAs) or through the future space-based gravitational wave observatory, {\it LISA} e.g. \cite{bara15}.

\section{Searching for periodic X-ray emission from galaxy centres}

We have carried out a systematic search for periodicities in all X-ray detections in the 4XMM-DR13 catalogue that have at least 100 EPIC counts. As indicated in Sec.~\ref{sec:intro}, these sources have lightcurves made in the automatic pipeline, for each observation. To search for periodic behaviour, we first selected the best quality detections, those that have a SUM\_FLAG equal to 0 (sources that are considered as good) and then barycentred the lightcurve data using the position of the source.  We retained only good time intervals, as background flares due to soft protons can result in false periodicity detections. If the filtering removed more than 75\% of the exposure, we did not search for a periodicity.  To search for periodic behaviour we used the accelerated search \citep{rans02} technique implemented in the HENDRICS python package \citep{hupp19,bach18b}.  To confirm the periods, we also implemented the Stingray/HENDRICS epoch folding \citep{leah83} and Z$_n^2$ \citep[for n=1 - simple sinusoidal modulation for rapidity][]{bucc83, bach21} techniques. We searched for modulation that was shorter than 0.2 of the observation duration, to allow at least five cycles to be detected in a single observation. We also ran our code on the background regions, to ensure that we were not detecting false periodicities from the background. This also helped to identify any possible period that was associated with background flares that were not completely excluded. We tested our pipeline on sources that were known to show periodic modulation, such as 2XMM J123103.2+110648 \cite{lin13}. Once this was shown to work, we ran it on the whole sample. We retained candidates that had at least a 4 $\sigma$ detection. 1355 detections showed pulsations, of which 345 were identified as detections within 3$\sigma$ of the centre of galaxies present in the catalogue GLADE+ \citep{daly22}. For the 1010 sources not apparently coincident with the centre of known galaxies, about one third are stars, where the star is not necessarily periodic, but a single strong burst may provide a false periodicity, in the same way as the background flare. About a fifth are associated with AGN, and a few tens are associated with X-ray binaries,  cataclysmic variables and ULXs. A similar number are associated with galaxies not in GLADE+, but about 300 sources are still to be identified. Concerning the sources concident with the centre of a GLADE+ galaxy, a dozen have $>$5$\sigma$ significant periodic variability. This is often not strictly sinusoidal, which is expected if the orbit shows ellipticity. The galaxies have estimated black hole masses of $\sim$10$^7$ - 10$^8$ M$_\odot$ except in one case, for a dwarf galaxy, where the black hole mass is estimated to be $\sim$1300 M$_\odot$. Supposing binaries with similar mass black holes in circular orbits implies typical separations of 0.002-0.01 mpc. Some of these systems may therefore be detectable with {\it LISA}. Work is still ongoing to validate these candidates, to exclude the possibility of red noise fluctuations from standard accretion processes being spuriously interpreted as periodic modulation.

\section{Finding X-ray transients in almost real-time with XMM-Newton}

As of 2025, almost real-time XMM-Newton transients that vary by at least a factor 5 in flux are being communicated via the link \url{http://flix.irap.omp.eu/stonks}. This has been achieved by carefully constructing a catalogue of X-ray detections made with the {\it XMM-Newton, Chandra, Swift, ROSAT}, and {\it eROSITA} observatories and using {\it XMM-Newton} upper limits. As the new {\it XMM-Newton} observations are reduced in the automatic pipeline \citep{quin24},  the fluxes are automatically compared with detections in the master catalogue, using a new functionality called STONKS. This is done for observations where a PI has given their prior accord for transient sources to be made public. Each automatic alert is manually screened to ensure its reliability, before being made public.  The alerts show both rising and declining flux and a number of nuclear transients have been identified. Here we present nuclear transients that have moved into the faint phase and can no longer be followed up. Ongoing analysis of currently bright transients will be the subject of future papers. 

\subsection{Tidal disruption events found with STONKS}

Here we detail possible candidate tidal disruption events found with STONKS as their flux declined towards quiescence and therefore are not followed-up with complementary observations.

\subsubsection{4XMM J071928.1+591102}

\begin{figure}[t]
\begin{center}    
{\includegraphics[width=240pt,height=20pc]{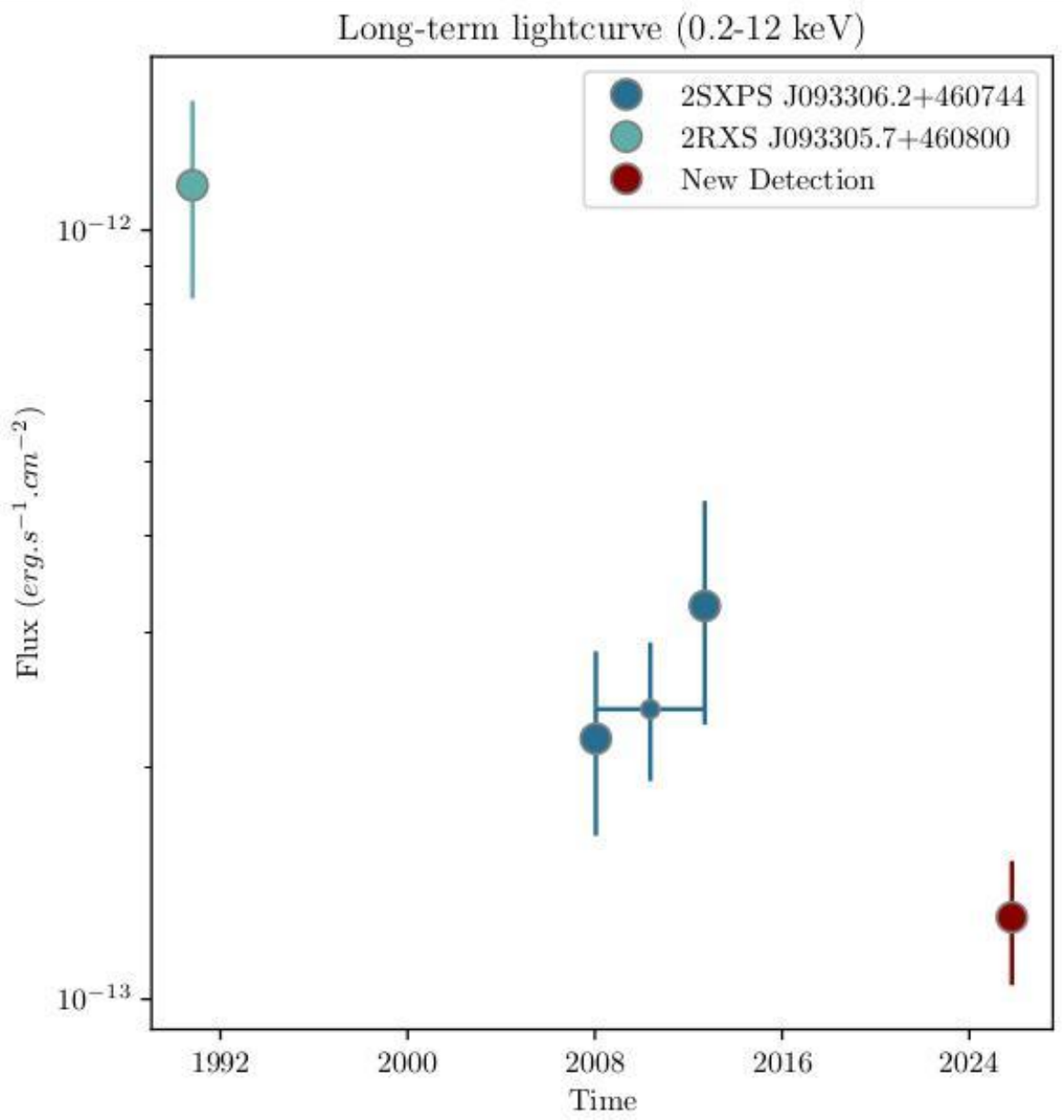}}
\caption{Long-term STONKS X-ray lightcurve of the transient J093306.2+460750.1\label{fig:J093306.2+460750.1}}
\end{center}
\end{figure}

4XMM J071928.1+591102 was first detected with ROSAT in 1991 at a flux level of $\sim$6.2 $\times$ 10$^{-13}$ erg cm$^{-2}$ s$^{-1}$, with a soft spectrum. It was detected again with XMM-Newton in September 2010 a factor three fainter and with a considerably hardened spectrum, well fitted with an absorbed (2.6 $\times$ 10$^{21}$ cm$^2$) power law with a photon index of 2.2. A possible low significance burst is seen in the lightcurve. Following a Swift observation a couple of years later, where the source continued to decline in flux, it was detected in October 2025 with a highly significant EPIC maximum likelihood of detection of 53, in a state similar, but considerably fainter, to the 2010 detection. This detection is coincident with the centre of the  WISE galaxy J071928.11+591102.6   situated at 1.67 Gpc \citep{daly22}. Estimating the black hole mass from the stellar mass in this latter catalogue ($\sim$7 $\times$ 10$^{10}$ M$_\odot$) using scaling relations \citep{rein15} implies a black hole of $\sim$7 $\times$ 10$^7$ M$_\odot$. At peak this source had an X-ray luminosity of  $\sim$2 $\times$ 10$^{44}$ erg s$^{-1}$ (0.2-12.0 keV) implying that the source was emitting at about 10\% Eddington during the ROSAT observations. The factor of almost 10 flux decline in 34 years (a very long decay time) could imply a TDE, but the sparse lightcurve makes it difficult to confirm the TDE nature.

\subsubsection{4XMM J081738.2+012402.9}

4XMM J081738.2+012402.9 was first detected with Swift in 2008, with a soft spectrum. 17 years later, XMM-Newton detected the source a factor $\sim$20 fainter, but highly significant with an EPIC maximum likelihood of detection of 82, in a harder state, reminiscent of the transition to a low-hard state seen in many of the X-ray TDEs described in Sec.~\ref{sec:transients}.  This detection is situated at 1.86" (less than 2$\sigma$) from the centre of the WISE galaxy WISEA J081738.15+012404.1. However, this maybe a neighbouring galaxy, with another similar, but slightly fainter source appearing to be the possible counterpart to the X-ray source. Again the very sparse lightcurve and faint nature of the source currently makes it difficult to confirm a TDE origin.

\subsubsection{4XMM J093220.2+460634.9}

4XMM J093220.2+460634.9 was first detected with ROSAT in 1991 with a very soft spectrum. 34 years later, XMM-Newton detected the source a factor $\sim$15 fainter, but highly significant with an EPIC maximum likelihood of detection of 188, in a hard power law state, reminiscent of the transition to a low-hard state seen in many of the X-ray TDEs described in Sec.~\ref{sec:transients}.  This detection is coincident with a very red and probably extended AllWISE source and could be a distant galaxy. Again the very sparse lightcurve and faint nature of the source currently makes it difficult to confirm a TDE origin.

\subsubsection{4XMM J093306.2+460750.1}

4XMM J093306.2+460750.1 was first detected with ROSAT in 1991, see Figure~\ref{fig:J093306.2+460750.1} with a very soft spectrum. Several Swift pointings were made 17 years later when the source was almost a factor ten fainter. Intriguingly, non-significant increased variability was seen during these observations, which were still very soft, and could be reminiscent of QPEs, but no significant flare could be seen in the Swift observation, due to the very limited number of counts. Eight years later, XMM-Newton detected the source a factor 2-3 fainter, but highly significant with a maximum likelihood of detection of 381, in a hard state, reminiscent of the transition to a low-hard state.  This detection is compatible (at 3$\sigma$) with 2MASX J09330606+4607516 a Seyfert 1 galaxy situated at 723 Mpc \cite{daly22}. This implies an X-ray luminosity of  $\sim$6 $\times$ 10$^{43}$ erg s$^{-1}$ (0.2-12.0 keV) at peak, very similar to peak luminosities of the TDEs described in Sec.~\ref{sec:transients}. Estimating the black hole mass from the stellar mass using scaling relations as above implies a black hole of $\sim$7 $\times$ 10$^7$ M$_\odot$, implying that the source was about a few percent of Eddington during the ROSAT observations. Alternative explanations for the emission include Seyfert I variability, which are known to show similar variability. The lack of observations between ROSAT and Swift make it difficult to conclude on the nature of the variability.

\section{Conclusions}\label{sec5}

We reviewed the current XMM-Newton catalogues available and demonstrated their use in finding nuclear transients, providing several examples of tidal disruption events and quasi-periodic eruption sources. We also showed preliminary work on a search for periodic variables in the XMM-Newton EPIC archival data, with the example of finding new massive black hole binaries. We described the STONKS pipeline that is now in the automatic reduction pipeline for XMM-Newton and the alert system in place to allow researchers to follow-up new and fading transients in almost real-time. We examined some fading sources detected with STONKS that may be newly identified TDE candidates, but which are too faint to validate fully. Upcoming papers will present new, bright TDEs discovered with the same method. The upcoming 5XMM catalogue, which will provide many new products compared to the 4XMM catalogues, including source classification, long-term variability, photometric redshifts, multi-wavelength counterparts, including OM sources and spectral fits, will provide new opportunities for identifying and studying nuclear transients. 


\section*{Acknowledgments}

This work was funded in part by the CNES \& the Horizon 2020 research and innovation programme under grant agreement n°101004168, the XMM2ATHENA project.










\bibliography{Wiley-ASNA}%

\end{document}